\documentclass[conference]{IEEEtran}
\IEEEoverridecommandlockouts
\usepackage{cite}
\usepackage{amsmath,amssymb,amsfonts}
\usepackage{algorithmic}
\usepackage{algorithm}
\usepackage{array}
\usepackage{graphicx}
\usepackage{booktabs}
\usepackage{textcomp}
\usepackage{footnote}
\usepackage{xcolor}
\usepackage[hidelinks,pdfpagemode=UseNone]{hyperref}
\usepackage[algo2e,linesnumbered,lined,boxed,commentsnumbered,ruled]{algorithm2e}
\def\BibTeX{{\rm B\kern-.05em{\sc i\kern-.025em b}\kern-.08em
    T\kern-.1667em\lower.7ex\hbox{E}\kern-.125emX}}

\makeatletter
\newcommand{\linebreakand}{%
  \end{@IEEEauthorhalign}
  \hfill\mbox{}\par
  \mbox{}\hfill\begin{@IEEEauthorhalign}
}
\makeatother

\def\authorname#1{{#1}\\}
\def\address#1{{\em #1}\\} 

\title{ConSinger: Efficient High-Fidelity Singing Voice Generation with Minimal Steps}

\author{
    \authorname{Yulin Song\textsuperscript{*}, 
    Guorui Sang\textsuperscript{*}, 
    Jing Yu,
    Chuangbai Xiao\textsuperscript{\textdagger}}
    \address{Beijing University of Technology}
    \{keylxiao, maluyelang\}@emails.bjut.edu.cn,
    \{jing.yu, cbxiao\}@bjut.edu.cn
     \thanks{\textsuperscript{*}Equal contribution.} \thanks{\textsuperscript{\textdagger}Corresponding author.}
}

\begin{document}

\maketitle

\begin{abstract}
Singing voice synthesis (SVS) system is expected to generate high-fidelity singing voice from given music scores (lyrics, duration and pitch). Recently, diffusion models have performed well in this field. However, sacrificing inference speed to exchange with high-quality sample generation limits its application scenarios. In order to obtain high quality synthetic singing voice more efficiently, we propose a singing voice synthesis method based on the consistency model, ConSinger, to achieve high-fidelity singing voice synthesis with minimal steps. The model is trained by applying consistency constraint and the generation quality is greatly improved at the expense of a small amount of inference speed. Our experiments show that ConSinger is highly competitive with the baseline model in terms of generation speed and quality. Audio samples are available at \href{https://keylxiao.github.io/consinger}{https://keylxiao.github.io/consinger}.
\end{abstract}


\section{Introduction}
\label{sec:intro}

SVS is designed to generate emotionally realistic human audio. Initial SVS Systems \cite{Macon1997ConcatenationBasedMV, kenmochi2007vocaloid} can automatically synthesize a singing voice by concatenating short waveform units selected from database. Two unwieldy problems with this architecture are that units do not always connected smoothly and even using large database can not avoid the limited flexibility of synthesis. Statistical parametric approach \cite{saino2006hmm, oura2010recent} based on hidden Markov models (HMM) have been proposed to solve the problems mentioned above. A HMM is trained to predict the singing voice waveform, enabling the use of less data to construct audio. Whereas such systems degrade the naturalness of the synthesized singing voice and lack of human-like variation due to over-smoothing. With the rapid development of deep learning, SVS system evolves into a two-stage generation: first, an acoustic model interprets the music score as the corresponding acoustic features (e.g., mel-spectrograms or World vocoder parameters \cite{morise2016world}); then, a vocoder is used to convert the generated features into audio waveforms. Several new-type SVS systems based on Generative Adversarial Network (GAN) \cite{chen2020hifisinger,xu22d_interspeech} and Denoising Diffusion Probabilistic Model (DDPM) \cite{liu2022diffsinger,cho2022mandarin} have been proposed to synthesize higher quality singing voice. However, these methods suffer from less stable training or slow inference speed. Achieving high speed inference while maintaining high quality sampling has become a challenging task.

Recently, consistency model \cite{Song2023ConsistencyM}, a model that balances high generation speed with satisfactory sampling quality, shows excellent performance in many fields, such as image generation \cite{Song2023ConsistencyM}, controllable music generation \cite{novack2024ditto} and video generation \cite{wang2023videolcm}. Both the distillation method and the isolated training method have been successfully applied in the Text-to-speech (TTS) field \cite{ye2023comospeech,li2024cm}, but the dependence on the teacher model and two training networks increase the training burden.

We propose ConSinger, a consistency model for SVS, which does not require a teacher model and only needs a single training network. ConSinger can be trained by optimizing consistency loss and generates mel-spectrogram approximatively matching the ground truth distribution with minimal steps. Inspired by the shallow diffusion mechanism \cite{liu2022diffsinger}, ConSinger can further improve the singing quality at the slight cost of generating speed. Based on this, we review the denoising generative model \cite{song2019generative,ho2020denoising,song2021scorebased} from another perspective and propose the final version: it can get better generation quality with the same generation speed. The evaluation results show that our method is highly competitive compared to baseline models in terms of generation speed and quality. The key contributions of this work can be summarized as follows:
\begin{itemize}
\item We proposed a SVS method based on the consistency model for generating high-fidelity singing voice in real time with minimal steps.
\item We perform multiple evaluations and analyses of the generated results, further comparing and relating them to the shallow diffusion mechanism proposed by \cite{liu2022diffsinger}.
\end{itemize}
\begin{figure*}[htb]\
\begin{minipage}[b]{0.48\linewidth}
    \centering
    \centerline{\includegraphics[width=7.5cm]{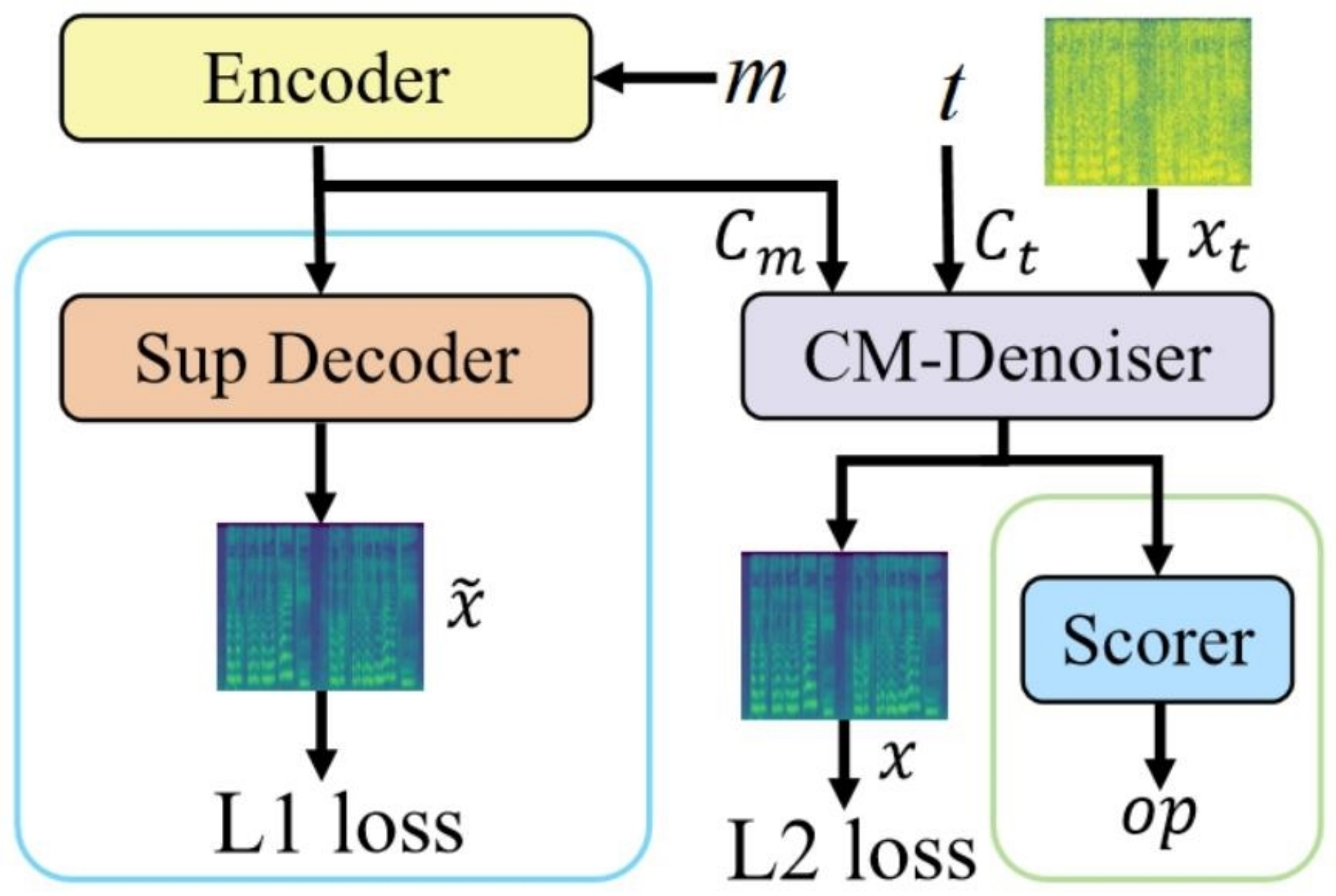}}
    \centerline{(a) the training pipeline of ConSinger.}
    \label{infer}
\end{minipage}
\hfill
\begin{minipage}[b]{0.48\linewidth}
    \centering
    \centerline{\includegraphics[width=8.5cm]{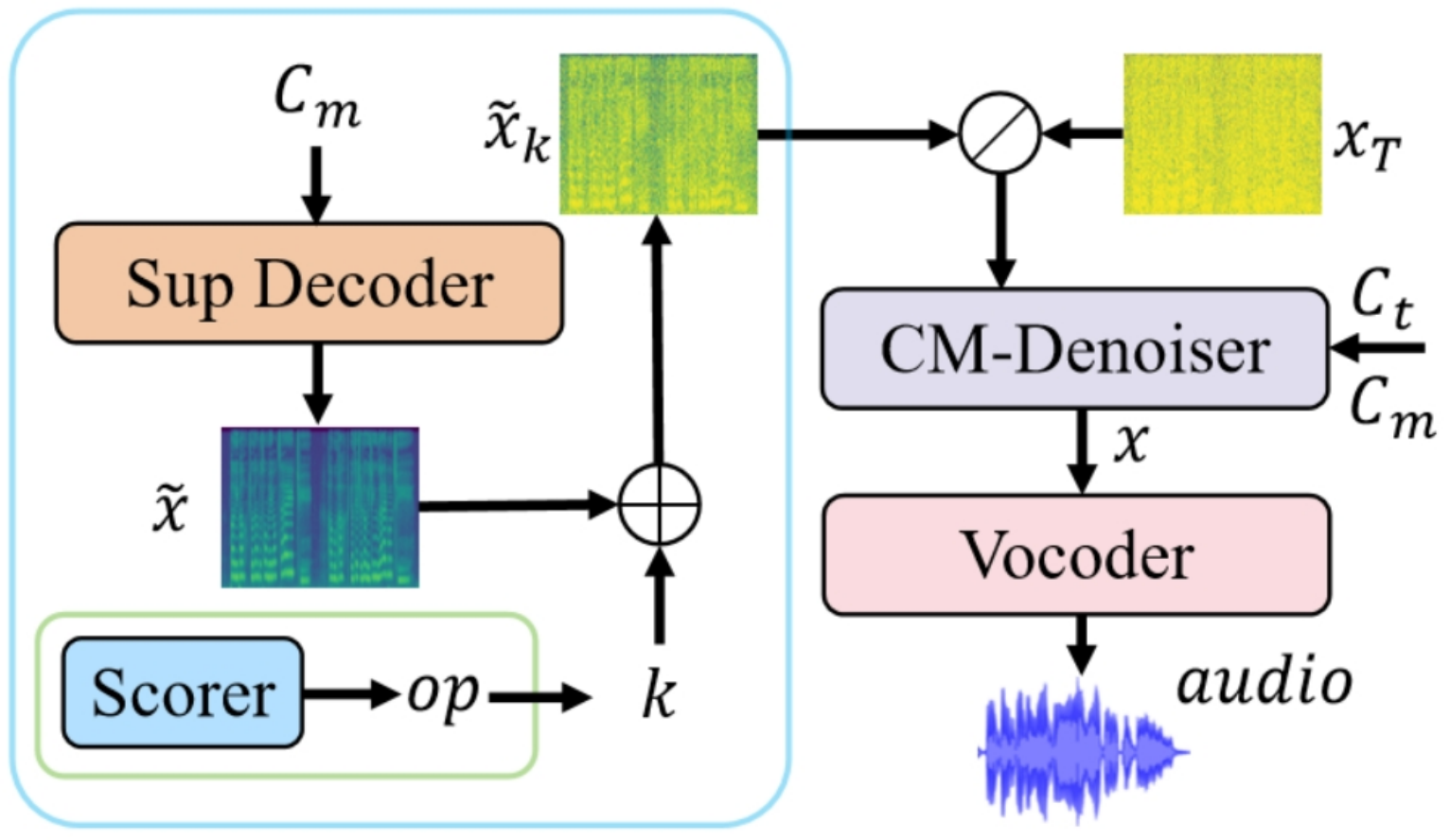}}
    \centerline{(b) the inference pipelins of ConSinger.}
    \label{train}
\end{minipage}
  \caption{The training and inference pipelines of ConSinger, the supplementary decoder and the scorer are included in the two light colored boxes. $m$ is the music score; t is the step number; $C_t$ and $C_m$ are their embedding information; $\tilde{x}$ is the mel-spectrogram generated by supplementary decoder; $x_t$ means the ground truth mel-spectrogram $x$ with $t$-level Gaussian noise; $op$ is the optimal point calculated by scorer. $k$ is the noise-add level calculated in place of $T$.}
  \label{fig:train&infer}
\end{figure*}
\section{Related Work}
\label{sec: RW}

To address the shortcomings of diffusion models\cite{Song2023ConsistencyM}, proposed a consistency character for \textit{probabilistic flow ordinary differential equation} (PF ODE) with the following conditions:
\begin{equation}
\begin{split}
    & f(\mathbf{x}_{\epsilon},\epsilon)=\mathbf{x}_{\epsilon},\\
    & f(\mathbf{x}_{t},t)=f(\mathbf{x}_{t'},t').
\end{split}
\label{loss condition}
\end{equation}
According to \cite{karras2022elucidating}, the starting point of the PF ODE is set to a small positive number $\epsilon$ for numerical stability. Then, we utilize a kind of skip connection to satisfy both conditions above:
\begin{equation}
\label{model function}
    f_{\theta}(\mathbf{x}_t,t) = c_{skip}(t)\mathbf{x}_t + c_{out}(t)F_{\theta}(\mathbf{x}_t,t).
\end{equation}
Here, $F_{\theta}(\cdot,\cdot)$ represents a neural network and $f_{\theta}(\cdot,\cdot)$ is the final output of models. $c_{skip}$ and $c_{out}$ are differentiable functions, while $t=\epsilon$, $c_{skip}=1$ and $c_{out}=0$, we can satisfy this by setting:
\begin{equation}
\label{cskipout}
    c_{skip}(t) = \frac{\sigma^{2}_{data}}{(t-\epsilon)^2+\sigma^{2}_{data}}, c_{out}(t) = \frac{\sigma_{data}(t-\epsilon)}{\sqrt{\sigma^{2}_{data}+t^2}},
\end{equation}
where $\sigma_{data}$ is a balance parameter, the points on the PF ODE trajectory can be sampled by:
\begin{equation}
\label{get xt}
    \mathbf{x}_{t} = \mathbf{x}_{\epsilon} + t_{n} \mathbf{z},
\end{equation}
where $\textbf{z}\sim\mathcal{N}(0,\textbf{I})$ and time step $t_{n}$ ,which can also be regarded as the noise level, is obtained as follows:
\begin{equation}
\label{sample t}
    t_{n} = \left[\epsilon
    ^{\frac{1}{\rho}}+\frac{n-1}{N-1}(T^{\frac{1}{\rho}}-\epsilon^{\frac{1}{\rho}})\right]^\rho.
\end{equation}
With the above as precondition, all points $p_{t}(\mathbf{x})$ along the sampling trajectory of PF ODE are directly associated with the original data distribution point $p_{0}(\mathbf{x})$, such that one-step generation can be achieved. In this study, we opt for isolated training method and propose enhancements to it.

\section{METHODOLOGY}
\label{sec:method}

The training and inference pipelines of ConSinger are depicted in Figure \ref{fig:train&infer}. In this Section, we first present the architecture of ConSinger; then we propose three versions of ConSinger.
\subsection{Model Architecture}
\label{model architectures}

\noindent \textbf{Encoder:}  We utilize the encoder structure in \cite{liu2022diffsinger}, which converts music score into score condition sequence $C_{m}$. Specifically, a lyrics encoder and $N$ feed-forward transformer blocks \cite{vaswani2017attention} turn phoneme ID into linguistic sequence. Pitch embedding sequence will be generated by pitch ID through a pitch encoder. Then, the duration predictor projects the linguistic sequence into the sequence in the mel-spectrogram domain. Finally, the encoder binds the linguistic sequence and pitch sequence into a sequence of music score conditions, denoted as $C_m$.

\noindent \textbf{Supplementary Decoder:}  We use a simple mel-spectrogram decoder as the supplementary decoder, which is consistent with decoders in \cite{blaauw2020sequence,ren2021fastspeech}. Specifically, the decoder is based on a feed-forward variant of the Transformer model \cite{vaswani2017attention, ren2019fastspeech}. Each layer is composed of a self-attention sublayer block and a convolutional sublayer block. Both sublayers are equipped with residual connections, layer normalization and dropout mechanism.

\noindent \textbf{CM-Denoiser:}  The CM-Denoiser is expected to restore the ground truth mel-spectrogram from Gaussian noise. The diffusion model is flexible in its network requirement, such that we have a wide array of alternatives available. Here we use a non-causal WaveNet \cite{vandenoord16_ssw,rethage2018wavenet,kong2021diffwave}.

\noindent \textbf{Scorer:}  The Scorer is designed to take a few reconstruction samples during training to compare with the reference ground truth samples in order to obtain the optimal denoising level $op$. We adopt the Fréchet Audio Distance (FAD) \cite{Kilgour2019FrchetAD} as the reference score. 

\noindent \textbf{Time Step Processing:}  We use sinusoidal position embedding to transform the time step $t$ into a continuous hidden condition $C_{t}$ \cite{vaswani2017attention}.

\noindent \textbf{Vocoder:}  In the final stage of inference, the model needs to utilize a vocoder to convert mel-spectrograms generated by CM-Denoiser into waveforms perceptible to humans.
\subsection{Initial Version}
\label{inital version}
In the training procedure, ConSinger obtains a $t$-level noise-add mel-spectrogram $\mathbf{x}_{t}$ and predicts the ground truth $\mathbf{x}$ based on $C_{t}$ and $C_{m}$. To satisfy both conditions of  Eq. (\ref{loss condition}), we define the loss function as:
\begin{equation}
\begin{split}
     &\mathcal{L}(\theta) = 
     \left|\left|\mathbf{x}-f_{\theta}(\mathbf{x}_{n},m,t_{n})\right|\right|^2.
\end{split}
\end{equation}

\noindent \textbf{Importance Sampler:} We designed a simple sampler to obtain the time step $t_{n}$ in Eq. (\ref{sample t}) more accurately. Concretely, a loss table $L(\cdot)$ records the cumulative average loss at each point on the trajectory and serves as a guide in the selection of sampling points. The formulation is given by $h_{n}=(1-\lambda)\frac{L(n)}{\sum^{N}_{i=2}L(i)}+\lambda.$ Here, $\lambda$ is the equilibrium parameter, which is used to adjust the model for random sampling and importance sampling. The model takes 10 samples from each sampling point before enabling the importance sampler to reduce the randomness.

The inference procedure involves sampling from a $T$-level Gaussian noise distribution $\mathcal{N}(0,T^2\mathbf{I})$ and utilize it to predict ground truth mel-spectrogram.
\subsection{Use Supplementary Decoder}
\label{use sup decoder}
Utilizing $T$ times standard Gaussian noise as the starting point for one-step restore poses a considerable challenge for the network. Inspired by the shallow diffusion mechanism \cite{liu2022diffsinger}, we can use a previous simple acoustic model to provide a large amount of prior knowledge for ConSinger. Given a data sample $\mathbf{x}$ and its corresponding $\tilde{\mathbf{x}}$ (generated by supplementary decoder), the conditional distributions of $\mathbf{x}$ and $\tilde{\mathbf{x}}_{t}$ are:
\begin{equation}
\begin{split}
    & q(\mathbf{x}_{t}\mid \mathbf{x})=\mathcal{N}(\mathbf{x}_{t};\mathbf{x},t^{2}\textbf{I}), \\
    & q(\tilde{\mathbf{x}}_{t}\mid \tilde{\mathbf{x}})=\mathcal{N}(\tilde{\mathbf{x}}_{t};\tilde{\mathbf{x}},t^{2}\textbf{I}).
\end{split}
\end{equation}
The KL-divergence between two Gaussian distributions is:
\begin{equation}
\begin{split}
        & D_{KL}(\mathcal{N}(\mathbf{x}_{t}) \left|\right|\mathcal{N}(\tilde{\mathbf{x}}_{t})) = \frac{1}{2}[ tr\left(\tilde{\Sigma}^{-1}\Sigma\right) + \\ 
        & (\tilde{\mu}-\mu)^{\top}\tilde{\Sigma}^{-1}(\tilde{\mu}-\mu) -d+\ln\left(\frac{\det\tilde{\Sigma}}{\det\Sigma}\right)] \\
        &=  \frac{{\left|\right|\mathbf{x}-\tilde{\mathbf{x}}\left|\right|}^{2}_{2}}{2t^{2}}.
\end{split}
\end{equation}
Here, $\tilde{\mu}$, $\mu$ are means; $\tilde{\Sigma}$ and $\Sigma$ are covariance matrices; d is the dimension. 
\begin{equation}
\label{calculate k}
\begin{split}
        &\mathrm{E}_{\mathbf{x} \in \mathcal{D}} \left[ D_{KL} \left( \mathcal{N}(\mathbf{x}_{k}) \parallel \mathcal{N}(\tilde{\mathbf{x}}_{k}) \right) \right] \\
        & = \mathrm{E}_{\mathbf{x} \in \mathcal{D}} \left[ \frac{1}{2k^{2}} \parallel \mathbf{x} - \tilde{\mathbf{x}} \parallel_{2}^{2} \right] \\
        &\le \mathrm{E}_{\mathbf{x} \in \mathcal{D}} \left[ D_{KL} \left( \mathcal{N}(\mathbf{x}_{T}) \parallel \mathcal{N}(0, T^{2} \mathbf{I}) \right) \right],
\end{split}
\end{equation}
which means using the mel-spectrogram with $k$-level noise as the restore point is at least no worse than Gaussian distribution $\mathcal{N}(0,T^2 \textbf{I})$. The value of $k$ is continuously optimized in the training phase.

\subsection{Use Scorer to Determine the Optimal Point}
\label{find optimal point}
During the training phase, we introduce a scorer to derive the optimal restore point $op$ on the trajectory, replacing the $k$ calculated in Section \ref{use sup decoder}.

\begin{figure}[htb]

\begin{minipage}[b]{.49\linewidth}
    \centering
    \centerline{\includegraphics[width=4.0cm]{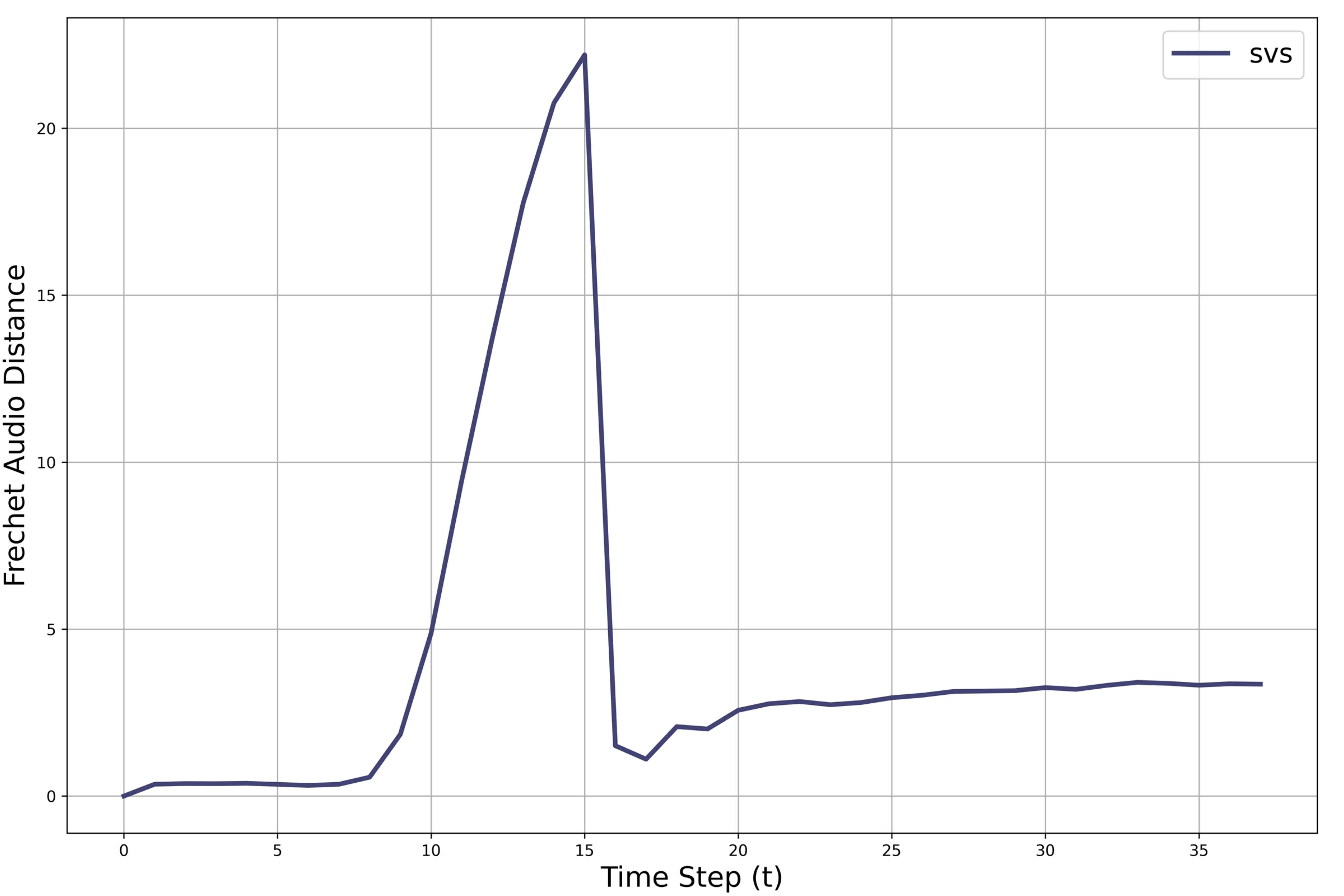}}
    \centerline{(a) FAD of different PF points.}
    \label{fad with gt}
\end{minipage}
  \hfill
  \begin{minipage}[b]{.49\linewidth}
    \centering
    \centerline{\includegraphics[width=4.0cm]{figs/t.pdf}}
    \centerline{(b) model parameters.}
    \label{param}
\end{minipage}
\caption{The subfigure (a) shows the mel-spectrogram quality restored from different noise level. The points on the x-axis with values of 0 and 1 represent GT and GT (mel+HiFi-GAN). The subfigure (b) shows the function curves of three important parameters of the model (Eq. (\ref{cskipout}) \& Eq. (\ref{sample t}), $\rho = 7$) and the time step $t$ (Normalized) in the consistency model.}
\label{fig:all fad}
\end{figure}

When performing ablation experiments, we found that the result quality restored by ConSinger (v2) is not simply linear with the level of noise (Figure \ref{fig:all fad}). We consider this is because: 1) ConSinger outputs a weighted combination of the input noise and the network output in Eq. (\ref{model function}); 2) in Eq. (\ref{cskipout}), $c_{skip}$ is transformed from 1 to 0, while $c_{out}$ is transformed from 0 to 0.5; 3) the time step, which can be viewed as the noise level, is quasi-exponential in Eq. (\ref{sample t}). 

We can interpret Figure \ref{fig:all fad} in a piecewise manner: 1) from 2 to 6, ConSinger gradually reduces the proportion of the input mel-spectrogram $\tilde{\mathbf{x}}_{t}$, which means the impact of the noise output by the supplementary decoder continuously reduces. The increase in the proportion of network output improves the quality of generation. This part is consistent with the results obtained from ablation experiments in \cite{liu2022diffsinger}: when the denoising ability of the model is too weak (too few denoising steps in DiffSinger, low mixing ratio of the network in ConSinger), the results still carry some unknown distribution of noise from the supplementary decoder; 2) from 7 to 15, the mixed Gaussian noise in $\tilde{\mathbf{x}}$ increases sharply, yet the model remains overly trusting of $\tilde{\mathbf{x}}_{t}$, leading to a significant reduction in the ability to reconstruction. 3) from 16 to 37, as the proportion of network output in the mixed result increases, the restore result of ConSinger will gradually become stable, but it cannot reach the same quality as the first part due to excessive noise in $\tilde{\mathbf{x}}_{t}$. 

\begin{algorithm}[t] \caption{The training Procedure of ConSinger(v3)}\label{algorithm}
\textbf{Input}: CM-Denoiser $f_{\theta}$; reconstruction point $k$; initial model parameter $\theta$; the train dataset $(\mathcal{X},\mathcal{M})$; scorer $S$; importance sampler $IS$;
\begin{algorithmic}[1] 
\STATE \textbf{repeat}
    \STATE \quad Sample $(x,m)$ from $(\mathcal{X},\mathcal{M})$;
    \STATE \quad $\textbf{z} \sim \mathcal{N}(0,\textbf{I})$;
    \STATE \quad $n\sim IS\left(\{2,...,k\}\right)$;
    \STATE \quad $x_{n} \gets x + t_{n}\textbf{z}$;
    \STATE \quad $
        \mathcal{L}(\theta)\gets \left|\left|x-f_{\theta}(x_{n},m,t_{n})\right|\right|^2$;
    \STATE \quad $op \gets S\left(x,f_{\theta}(x_{n},m,t_{n})\right)$;
    \STATE \quad $IS \gets \mathcal{L}(\theta)$;
    \STATE \quad $\theta \gets \theta - \bigtriangledown_{\theta}\mathcal{L}(\theta)$;
\STATE \textbf{until} \textit{convergence;}
\end{algorithmic}
\end{algorithm}

\section{Experiments}
\subsection{Experimental Setup}
\noindent \textbf{Dataset:} We conduct our SVS experiments on PopCS dataset \cite{liu2022diffsinger}, which contains 117 Chinese pop songs (total around 5.89 hours with lyrics) collected from a qualified female vocalist, all recorded at 24kHz. Following \cite{blaauw2020sequence, ren2020deepsinger, liu2022diffsinger}, $F_{0}$ (fundamental frequency) of the original waveform is extracted as pitch information during testing. We randomly select clips from two songs for testing. 

\noindent \textbf{Inplementation Details:} We set the hop size and frame size to 128 and 512. The mel-spectrograms are linearly normalized to between [-1, 1], and the number of mel bins is 80; We set $\epsilon = 0.02$, $T = 80$, $\sigma_{data} = 0.5$ and $\rho = 7$, following \cite{Song2023ConsistencyM}. The number of points on the PF ODE trajectory is adjusted to 50. $\lambda$ in importance sampler is 0.05. We calculate $k = 37$ in Eq. (\ref{calculate k}); Other structural parameters of the model are consistent with \cite{liu2022diffsinger}.

\noindent \textbf{Evaluation Metrics:}  We use both subjective and objective assessments to evaluate the speed of model inference (NFE \& RTF) and the quality of generated samples (FAD \& MOS): 1) number of function evaluations (\textbf{NFE}) represents the total number of times the network is evaluated in the form of $x+y$, $x$ denotes the number of times the supplementary decoder was used, and $y$ represents the number of times the CM-denoiser was used; 2) real-time factor (\textbf{RTF}) is defined as the ratio between the total time the system takes to synthesize a given amount of audio and the duration of the audio; 3) Fréchet Audio Distance (\textbf{FAD}) \cite{Kilgour2019FrchetAD}. Here we use the same method as \cite{zhang2023amphion}; 4) mean opinion score (\textbf{MOS}) \cite{chu2006objective} is used to measure the subjective perceived quality of the synthesized song. In this work, 20 qualified listeners are asked to make judgments on the synthesized audio on a scale of 0 to 5.

\noindent \textbf{Training and Inference:} We train the encoder and the supplementary decoder for 160k steps. Then, we train CM-Denoiser 280k steps. The scorer update the optimal point every 2k steps. We employ a pre-trained HiFi-GAN \cite{kong2020hifi} for converting the predicted mel-spectrograms to audios.

\subsection{Evaluation Results}

We compared ConSinger with the following models: 1) GT, the ground truth singing voice; 2) GT(mel+HiFi-GAN), which converts reference audio to mel-spectorgrams and then restores it to audio using HiFi-GAN vocoder; 3) FFTSinger \cite{blaauw2020sequence} uses a supplementary decoder to generate mel-spectrograms and convert it to audio; 4) DiffSinger \cite{liu2022diffsinger}, a SVS system trained with diffusion models. 
\begin{table}[htbp]
    \caption{Evaluation Results on PopCS for SVS.}
    \small
    \centering
    \begin{tabular}{p{0.19\textwidth}cccc}
    \toprule
    method & NFE & RTF$\downarrow$ & FAD$\downarrow$ & MOS$\uparrow$ \\
    \midrule
    GT & - & - & - & 4.44\\
    GT(mel+HiFi-GAN) & - & - & 0.3542 & 4.33\\
    \midrule
    FFTSinger & 1+0 &\textbf{0.0149} & 1.9319 & 3.26\\
    DiffSinger & 1+51 & 0.0519 & 0.8779 & 3.81\\
    \midrule
    ConSinger(v1) & 0+1 &0.0171 & 3.5254 & 2.72\\
    ConSinger(v2) & 1+1 &0.0188 & 3.3510 & 2.96\\
    ConSinger(v3) & 1+1 &0.0183 & \textbf{0.3191} & \textbf{3.88}\\
    \bottomrule
    \end{tabular}
    \label{tab:svs}
\end{table}

From Table \ref{tab:svs}, ConSinger (v3) has similar generation speed with FFTSinger, but greatly improves the generation quality, which achieves optimal performance on objective and subjective quality indicators. At the same time, ConSinger (v3) greatly improves the generation quality (+1.16 MOS) at a 10\% reduction in the generation speed (RTF 0.0171 vs. 0.0188) by using the scorer, demonstrating the effectiveness of our approach.

\subsection{Ablation Studies}

We performed ablation experiments on ConSinger(v3) to further demonstrate the effectiveness of our approach. Concretely, We used Comparative Mean Opinion Score (CMOS) and Comparative Fréchet Audio Distance (CFAD) for evaluation. In Table \ref{tab:ablation}, $IS$ means using the importance sampler. $noise$ indicates whether consistency constraint is applied (Eq. \ref{model function}-\ref{sample t}), that is, whether to directly input the output of the supplementary decoder into the CM-denoiser. Note that both the importance sampler and $\rho$ are applied only in the consistency model. The results show that both the importance sampler and the use of consistency constraints contribute to improving the audio quality. Using other $\rho$ also results in a decrease in the quality.

\begin{table}[htbp]
    \caption{Ablation studies on ConSinger(v3).}
    \small
    \centering
    \begin{tabular}{c|c|c|c|c|c}
    \toprule
    No. & $IS$ & $noise$ & $\rho$ &CMOS & CFAD \\
    \hline
    1 & $\checkmark$ & $\checkmark$ & 7 & 0.00 & 0.000\\
    \midrule
    2 & $\times$ & $\checkmark$ & 7 & -0.06 & -0.008\\
    \midrule
    3 & $\times$ & $\times$ & - & -0.04 & -0.051\\
    \midrule
    4 & $\checkmark$ & $\checkmark$ & 4 & -0.23 & -0.011\\
    \bottomrule
    \end{tabular}
    \label{tab:ablation}
\end{table}

Same as DiffSinger, ConSinger (v2 \& v3) denoise prior knowledge with noise to get better results. That is, the supplementary decoder provides the mel-spectrogram skeleton for the model, the addition of a small amount of noise on this basis helps the denoiser to perform a more elaborate carving.
\section{Conclusion}
In this work, we introduced ConSinger, a SVS method based on the consistency model. In order to improve the generation quality and minimize the loss of inference time, we proposed the second version, which restores mel-spectrograms from intermediate point on the PF ODE trajectory rather than the boundary point. Inspired by the ablation experiments performed on this basis, we presented the final version of ConSinger and correlated the experiment results with the ablation experiment in DiffSinger. Our experiments show that DiffSinger does not fully exploit the performance of the network by predicting and denoising a small amount of noise at a time, and therefore it consumes abundant inference time. 

Although the generation speed is satisfactory, ConSinger requires a supplementary decoder to generate high quality audio, which burdens the training process. The experiment results indicate that there is still room for improvement in the generative quality of ConSinger. And in the training stage, the goal of scorer optimization is FAD, which may differ from the optimal results guided by the MOS test. We consider the limitations mentioned above as future work.
\section{Acknowledgment}

We thank participants of the listening test for the valuable evaluations.

\bibliographystyle{IEEEtran}
\bibliography{consinger}

\end{document}